\documentclass[aps,onecolumn]{revtex4}

\newcommand{\g}{j}
\newcommand{\kk}{\Gamma}
\newcommand{\wn}{\tau}
\newcommand{\F}{{\tilde F}}

\newcommand{\wt}{Q}
\newcommand{\evo}{population annealing }
\usepackage{amsmath,amssymb,graphicx}
\usepackage{algorithmic}
\begin{document}
\title{Population Annealing with Weighted Averages: A Monte Carlo Method for Rough Free Energy Landscapes}  
\author{J.~Machta}
\email{machta@physics.umass.edu}
\affiliation{Department of Physics, University of Massachusetts,
Amherst, Massachusetts 01003 USA}
\affiliation{Santa Fe Institute, 1399 Hyde Park Road, Santa Fe, New Mexico 87501 USA}

\begin{abstract}
The population annealing algorithm introduced by Hukushima and Iba is described.  Population annealing combines simulated annealing and Boltzmann weighted differential reproduction within a population of replicas to sample equilibrium states. Population annealing gives direct access to the free energy.   It is shown that unbiased measurements of observables can be obtained by weighted averages over many runs with weight factors related to the free energy estimate from the run.  Population annealing is well suited to parallelization and may be a useful alternative to parallel tempering for systems with rough free energy landscapes such as spin glasses. The method is demonstrated for spin glasses.
\end{abstract}

\maketitle
\section{Introduction}
One of the most difficult challenges in computational statistical physics is sampling the low temperature equilibrium states of systems with complex free energy landscapes such as occur, for example, in spin glasses, configurational glasses and biomolecules.    Standard Markov chain Monte Carlo (MCMC) methods such as the Metropolis algorithm become stuck in local minima and are unable to correctly sample the equilibrium distribution.    Attempts to solve this difficulty generally involve simulating a broadened distribution that smooths the free energy landscape.  One of the most widely used of these methods is parallel tempering or replica exchange Monte Carlo \cite{SwWa86,Geyer91,HuNe96}. In parallel tempering, many replicas of the system are simultaneously simulated using a standard MCMC method with each replica at a different temperature.  The sequence of temperatures  spans a range from high temperatures where the free energy landscape is smooth to the low temperatures of interest where it is rough.  Replica exchanges are allowed that  permit replicas to move between low and high temperatures.  The hope is that visits to high temperatures allow more rapid mixing of the Markov chain and equilibration between different minima.

Closely related to the problem of sampling equilibrium is the problem of finding ground states of systems with complex energy landscapes.  In computer science the same question arises for NP-hard combinatorial optimization problems. One generic method for finding ground states or at least low lying states is simulated annealing~\cite{KiGeVe83}.  In simulated annealing the system is subject to an equilibrating MCMC procedure at a sequence of temperatures.  Following this `annealing schedule' the system is gradually cooled through the sequence of temperatures from a high temperature where the free energy landscape is smooth to a low temperature where it is rough.  The hope is that the system will explore many minima while the temperature is high and settle into the deepest minima as the temperature is gradually lowered.

The population annealing algorithm introduced by Hukushima and Iba~\cite{HuIb03}  is based on simulated annealing~\cite{KiGeVe83} but also shares features with parallel tempering, histogram re-weighting~\cite{FeSw} and diffusion Monte Carlo~\cite{Anderson75}.    Like simulated annealing, it is a single pass algorithm with an annealing schedule. A population of replicas of the system is simultaneously cooled following the annealing schedule. Unlike simulated annealing, the population is maintained in equilibrium throughout the annealing process and an equilibrium sample is generated at every temperature.  Equilibrium is maintained by differential reproduction (resampling) of replicas according to their relative Boltzmann weights.  A useful by-product of the calculation of relative Boltzmann weights is direct access to the equilibrium free energy at every temperature. Methods for extracting free energy differences from population annealing are related to Jarzinski's equality~\cite{Jarz97}.  Population annealing is an example of a sequential Monte Carlo scheme \cite{DoFrGo01} and is related to nested sampling methods~\cite{Skilling06}.

In~\cite{HuIb03}, Hukushima and Iba compare population annealing to parallel tempering and show that they produce results with comparable efficiency.  However, they note that \evo suffers biases for small population size.  The main contribution of this work is to show that these biases, together with statistical errors, can be made arbitrarily small by using appropriate weighted averages over many independent runs of the algorithm.  The weight factor is obtained from the free energy estimator from each run and the variance of this estimator serves as a useful diagnostic for the algorithm.  As proof of principle of the method, we apply it to 1D and 3D spin glasses. 

In the next section of the paper population annealing is described.  In section \ref{sec:weight} the weighted averaging procedure is introduced.  Section \ref{sec:sg} describes simulations of the Ising spin glass and the paper closes with a discussion. 

\section{The Algorithm}
We now describe our implementation of the algorithm in detail.  We start with a population of $R_0$ replicas of the system.  For disordered spin systems, each replica has the same set of couplings.
Each replica is initialized at infinite temperature, $\beta=0$, thus each replica is in an independent microstate chosen at random with equal probability from among the $\Omega=\sum_\gamma 1$ microstates of the system.  For Ising systems with $N$ spins, the microstates $\gamma$ are described by $N$ binary variables and $\Omega=2^N$.   Without loss of generality, suppose that the average energy of the system at infinite temperature is zero, $ \langle E_\gamma \rangle_{\beta=0}=0$ as is the case for the Ising model.  The algorithm can also be used with a finite initial temperature but then absolute measurements of the free energy are not available.

The temperature of the set of replicas is now lowered while keeping the ensemble in equilibrium.  In order to accomplish this, the population is resampled---some replicas are reproduced and others are eliminated with the expected number of replicas held fixed.  Suppose we have a population of $\tilde{R}_\beta$ replicas chosen from an equilibrium ensemble at temperature $1/\beta$ and we want to lower the temperature to $1/\beta^\prime$ with $\beta^\prime > \beta$.  For a replica  $\g$, with energy $E_\g$ the ratio of the statistical weight at $\beta$ and $\beta^\prime$ is $\exp\left[ -(\beta^\prime-\beta)E_\g\right]$.
This reweighting factor is typically large so that a normalization is needed to keep the population size reasonable. We compute normalized  weights $\wn_\g(\beta,\beta^\prime)$ whose sum over the ensemble is $\tilde{R}_\beta$, 
\begin{equation}
\label{eq:wn}
\wn_\g(\beta,\beta^\prime)=\frac{\exp\left[-(\beta^\prime-\beta)E_{\g}\right]}{\wt(\beta,\beta^\prime)}
\end{equation}
where $\wt$ is the normalization given by
\begin{equation}
\label{eq:Q}
\wt(\beta,\beta^\prime)=\frac{\sum_{j=1}^{\tilde{R}_\beta} \exp\left[-(\beta^\prime-\beta)E_{\g}\right]}{\tilde{R}_\beta} .
\end{equation}

The new population of replicas is generated by resampling the original population.  The number of copies of state $\g$ in the new population  is ${\cal N}\left[(R_{\beta^\prime}/\tilde{R}_\beta)\wn_{\g}(\beta,\beta^\prime)\right]$ where ${\cal N}\left[a\right]$ is a Poisson random variate with expected value $a$.   If  ${\cal N}\left[(R_{\beta^\prime}/\tilde{R}_\beta)\wn_{\g}(\beta,\beta^\prime)\right]=0$, the configuration $\g$ is eliminated.  The size, $\tilde{R}_{\beta^\prime}$ of the population after the temperature step has expectation $R_{\beta^\prime}$.  In our implementation, the size of the population is variable but stays close to the set of target values $\{R_\beta\}$ in contrast to the fixed population method introduced in~\cite{HuIb03}.  

Of course, the new population is now more correlated than before since several replicas may be in  the same microstate.  Furthermore, to the extent that the energy distributions at the two temperatures do not overlap, the equilibrium distribution may no longer be adequately sampled at the lower temperature.  In order to mitigate both of these difficulties, all replicas are now subject to an equilibrating MCMC process at the new temperature.

The algorithm thus consists of a sequence of temperature steps with differential reproduction within the population according to the relative Boltzmann weights followed by additional equilibration at the new temperature using a standard MCMC procedure.  The ordered sequence of $K+1$ inverse temperatures, $\beta_0>\beta_1>\beta_2 \ldots >\beta_{K}=0$ determines the annealing schedule.  The population of $\tilde{R}_{\beta_k}$ replicas is equilibrated at $\beta_{k}$ using $\theta_k$ sweeps of the MCMC process.  The temperature is now lowered to $\beta_{k-1}$ with differential reproduction as described above followed by $\theta_{k-1}$ MCMC sweeps.  At each temperature, observables are measured after the MCMC sweeps and $\ln \wt$ is recorded.  In the present implementation, observables are measured {\it after} resampling so that no weight factors are needed in taking averages.  The temperature is lowered until $\beta_0$ is reached.  In the simplest version of the algorithm the target population sizes and number of MCMC sweeps are independent of temperature, $R_\beta = R$ and $\theta_k=\theta$ for all $\beta$.

Either the MCMC process or the differential reproduction steps are in principle sufficient to produce equilibrium at the new temperature though neither one individually is efficient. However, the combined processes are substantially more efficient than either alone.

One advantage of the algorithm is that it gives direct access to the free energy at each temperature.  The normalization factor $\wt(\beta,\beta^\prime)$ is an estimator of the ratio of the partition functions at the two temperatures,
\begin{eqnarray}
\label{nonlinear}
\frac{Z(\beta^\prime)}{Z(\beta)} &=& \frac{\sum_\gamma e^{-\beta^\prime E_\gamma}}{Z(\beta)}\nonumber\\
&=& \sum_\gamma e^{-(\beta^\prime-\beta) E_\gamma} ( \frac{e^{-\beta E_\gamma}}{Z(\beta)} )
\nonumber\\
&=& \langle e^{-(\beta^\prime-\beta) E_\gamma} \rangle_\beta
\nonumber\\
&\approx&  \frac{1}{\tilde{R}_\beta} \sum_{j=1}^{\tilde{R}_\beta} e^{-(\beta^\prime-\beta) E_{\g}}=\wt(\beta,\beta^\prime) .
\end{eqnarray}
Thus, the estimator of the free energy, $\F$ at each simulated temperature $1/\beta_k$ is,
\begin{equation}
\label{eq:sumQ}
-\beta_k \F(\beta_k) = \sum_{\ell=K}^{k+1} \ln \wt(\beta_{\ell},\beta_{\ell-1}) + \ln \Omega .
\end{equation}
This section concludes with pseudocode for  the algorithm:

\begin{center}
 {\bf Population Annealing}
 \end{center}
\algsetup{indent=2em}
\begin{algorithmic}
\STATE Initialize $R_0$ replicas at $\beta=0$
\FOR {$k=K$ {\bf to} 1 {\bf step} $-1$}
\STATE Compute the partition function ratio  $\wt(\beta_k,\beta_{k-1})$
\COMMENT {Eq.\ (\ref{eq:Q})}
\FORALL {$j \leq \tilde{R}_{\beta_{k}}$}
\STATE Compute the relative weight $\wn_j(\beta_k,\beta_{k-1})$
\COMMENT {Eq. (\ref{eq:wn})}
\STATE Resample: make ${\cal N}\left[(R_{\beta_{k-1}}/\tilde{R}_{\beta_k})\wn_{j}(\beta_k,\beta_{k-1})\right]$ copies of replica $j$\\
\COMMENT {${\cal N}(a)$ is a Poisson random variate with mean $a$}
\ENDFOR
\STATE Compute the population size $\tilde{R}_{\beta_{k-1}}$ 
\FORALL {$j \leq \tilde{R}_{\beta_{k-1}}$}
\STATE Equilibrate replica $j$ for $\theta_{k-1}$ Monte Carlo sweeps
\ENDFOR
\STATE Compute observables and the free energy at $\beta_{k-1}$
\COMMENT {Eq.\  \ref{eq:sumQ}}
\ENDFOR
\end{algorithmic}

\section{Weighted Averages}
\label{sec:weight}
Unless the population size, number of temperature steps and number of MCMC sweeps/step are  sufficiently large, the result of a single run of population annealing is significantly biased because the equilibrium distribution is not fully sampled. Simple averaging over an ensemble of independent runs does not reduce this bias. However, an appropriate  {\it weighted average} over an ensemble of independent runs effectively reduces both statistical errors and biases.  The free energy estimator from a given run of the algorithm represents the weight that should be assigned to the observations made during that run.  In particular, suppose that an observable $A$ is estimated in run $r$ to be $\tilde{A}_r(\beta)$ at temperature $1/\beta$.  
An unbiased estimate of $\overline{A}(\beta)$ from the entire simulation is the weighted average over the ensemble of $M$ independent runs of population annealing,
\begin{equation}
\label{eq:weight}
\overline{A}(\beta) = \sum_{r=1}^M \tilde{A}_r(\beta) \omega_r(\beta)
\end{equation}
where the weight for run  $r$ is given by
\begin{equation}
\label{eq:omega}
\omega_r(\beta)  = \frac{e^{-\beta \F_r(\beta) } }{\sum_{r=1}^M  e^{-\beta \F_r(\beta)}}
\end{equation}
and $\F_r(\beta)$ is the free energy estimator (\ref{eq:sumQ}) for the $r^{\rm th}$ run of the algorithm at temperature $1/\beta$. The weight factors $e^{-\beta \F_r(\beta)}$ are needed because they represent the total weight at that temperature that would have been associated with $r^{\rm th}$ run if the normalization factor in (\ref{eq:wn}) had not been used to keep the population size near the target size.  Runs with different but fixed target populations sizes $R^{(r)}$ can be combined using (\ref{eq:weight}) with $e^{-\beta \F_r(\beta)}$ replaced by $R^{(r)} e^{-\beta \F_r(\beta)}$ in both the numerator and denominator of (\ref{eq:omega}).
  
The free energy appears to require a more complicated weighting formula because it is the sum of terms from all temperatures. However, as shown below it may be expressed as a simple average. The quantity that is directly averaged at each temperature is $\wt$, not $\ln \wt$, so the logarithm must be taken after the weighted average.  The unbiased estimate of the free energy $\overline{F}(\beta_k)$ is given by,
\begin{eqnarray}
\label{eq:bestf}
\nonumber
-\beta_k \overline{F}(\beta_k) &=& \sum_{\ell=K}^{k+1} \ln \left[ \sum_{r=1}^M\wt_r(\beta_{\ell},\beta_{\ell-1}) \omega_r(\beta_\ell)\right] + \ln \Omega \\
\nonumber
&=&\sum_{\ell=K}^{k+1} \ln \left[ \frac{\sum_{r=1}^M \exp(-\beta_{\ell-1} \F(\beta_{\ell-1}))}{\sum_{r=1}^M \exp(-\beta_{\ell} \F(\beta_{\ell}))}\right]\\
&=& \ln \left[\frac{1}{M} \sum_{r=1}^M \exp(-\beta_{k} \F(\beta_{k}))\right]
 .
\end{eqnarray}
The inner summation indexed by $r$ is over the ensemble of independent runs while the outer summation indexed by $\ell$ is over the temperature steps between the highest temperature $\ell=K$ and the temperature of interest $\ell=k$.  The second expression follows from the first via (\ref{eq:sumQ}) and (\ref{eq:omega}).  The last expression shows that $\exp(-\beta \F(\beta))$ may be directly averaged.

Errors in the weighted averages can be obtained by resampling.  Here we use the bootstraps method~\cite{NeBa99} where the error is the standard deviation of a large number ($\sim 10^2$)  of resampled ensembles, each of size $M$.

Statistical and systematic errors in the method are determined by the probability distribution of 
the dimensionless free energy $-\beta \F$.  If this distribution has a variance that is much smaller than unity and tails that decay faster than exponentially then the important weight factors  (\ref{eq:omega}) do not vary much and the tails of the distribution are unimportant.  On the other hand, if the variance 
is larger than one, $\overline{A}$ is controlled by the upper tail of the distribution and the ensemble size $M$ must be large to reduce errors. 
In the case that $-\beta \F$ is normally distributed, the typical value of $-\beta \F$ that dominates the sum in (\ref{eq:weight}) is one variance (not one standard deviation!) above the mean.  The variance of $-\beta \F$ is thus a criteria for the performance of the algorithm.  If it is smaller than about 0.5 then a modest ensemble of independent runs will be sufficient to yield accurate results. Optimizing \evo should be guided by minimizing the variance of the free energy estimator for a fixed amount of computational effort.  Of course, it is always possible that the actual $-\beta \F$ distribution has significant weight far above its mean because of important low energy states that have never been explored.  This diabolical situation would similarly fool equilibration tests for parallel tempering~\cite{Mac09a}.

Population annealing using $M$ independent runs with population size $R$ is less accurate than a single run with population size $MR$. If ${\rm Var}(\beta \F)$ is small, the loss of accuracy is small and vice versa.  Nonetheless, an ensemble of independent runs has several important advantages.  First, independent runs with moderate $R$ require only a single processor or simple parallelism without communication between processors. Large values of $R$ are only practical with true parallelism involving communication between processor.  Second, the measurement of the variance of $\beta \F$ over the ensemble provides a diagnostic for the method.  Third, error bars are simply obtained from an ensemble using resampling.  

\section{Application to Ising Spin Glasses}
\label{sec:sg}
To illustrate the effectiveness of \evo for high precision simulations we apply it to the Ising spin glass in one and three dimensions.  
Ising spin glasses are notoriously difficult to equilibrate at low temperature.  
The Ising spin glass is defined by the energy
\begin{equation}
E=\sum_{( i,j )} J_{i,j} \sigma_i \sigma_j
\end{equation}
where $\sigma_i=\pm 1$ is the Ising spin at site $i$ and the summation is over the nearest neighbor bonds $( i,j )$ of the lattice.  We consider the simple cubic lattice (3D) with periodic boundary conditions and a periodic chain of spins (1D).  The couplings $J_{i,j}$ are quenched random variables chosen independently for each bond from a Gaussian distribution with mean zero and variance one.  The 3D Ising spin glass with Gaussian couplings has a continuous phase transition at  transition temperature  $T_c=1.052$ \cite{KaKoYo06}.   The low temperature phase of this model is still the subject of controversy. On the other hand, the 1D Ising spin glass is a trivial model.  Nonetheless, it has a rough free energy landscape and is difficult to equilibrate using generic Monte Carlo methods including parallel tempering~\cite{KaYo03,KaKoLiJuHa05}.  The 1D spin glass has the advantage that the Monte Carlo results can be compared with exact transfer matrix calculations.  

For the 1D spin glass we simulated 18 disorder realizations for a chain with $N=256$ spins using population annealing.  We report results for the lowest temperature, $1/\beta_0=0.2$.  We used a mean population size $R=1000$ with $K+1=100$ temperatures and $\theta=50$ Metropolis sweeps at each temperature.  Of the 99 temperature steps, 19 are equally spaced in inverse temperature 
between $\beta=0$ and $\beta=0.5$ while the remaining 80 are equally spaced in temperature.  Each run used $5\times 10^6$ Metropolis sweeps and for each realization the ensemble consisted of $M= 200$ independent runs.  

\begin{figure}[t]
\includegraphics[width=0.45\textwidth]{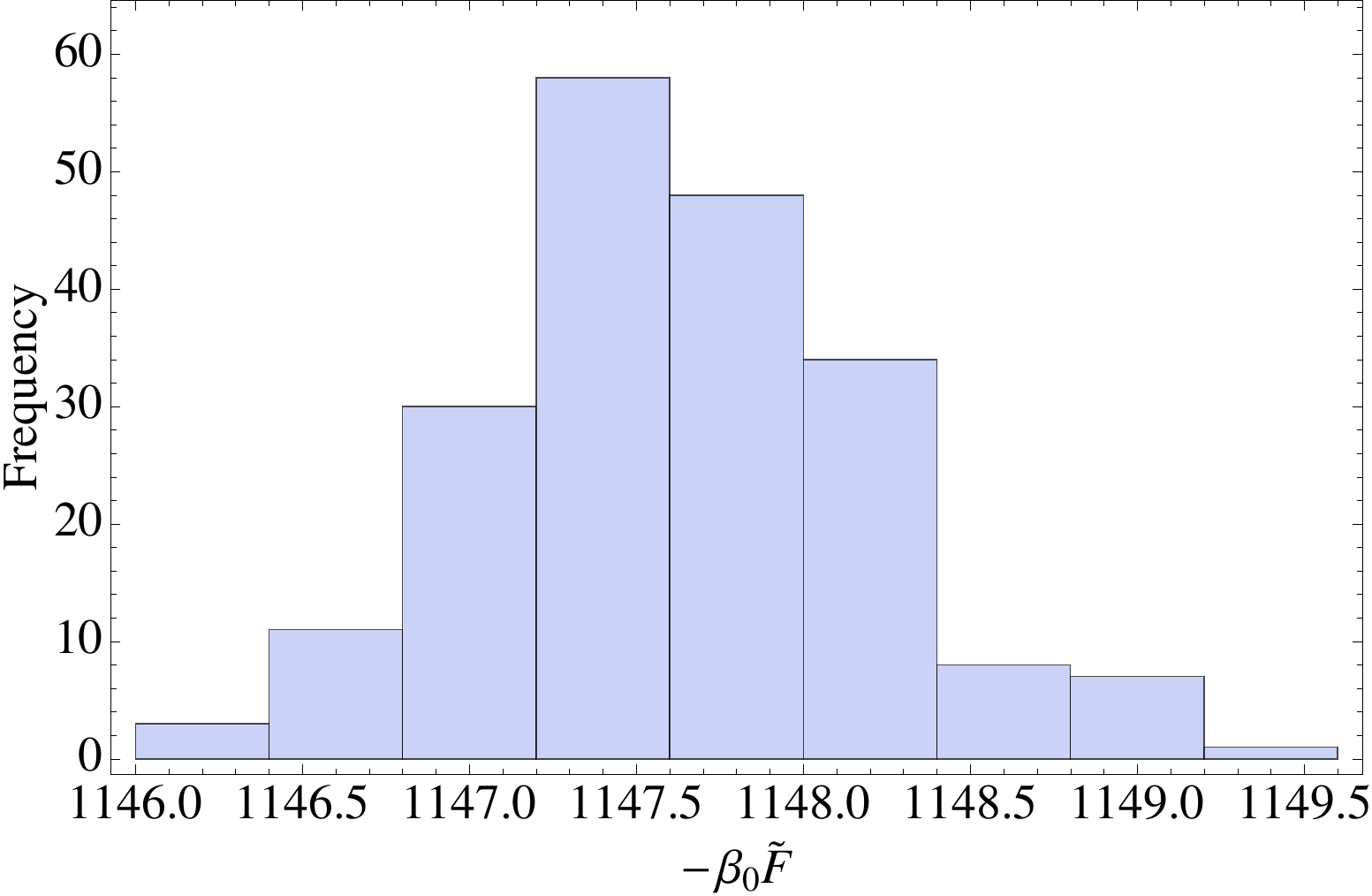}
\caption{The histogram of the dimensionless free energy, $-\beta_0 \F$ for population annealing applied to a single realization of a 1D Ising spin glass.   The exact value for this realization is $-\beta_0 F^{\rm exact}=1147.87$.}
\label{fig:free}
\end{figure}

First we report results for a single 1D disorder realization at the lowest temperature, $1/\beta_0=5$.
Figure \ref{fig:free} shows the histogram of the dimensionless free energy $-\beta_0 \F$.   
The variance of $-\beta_0 \F$ is 0.34 and the distribution does not appear to have fat tails.
Therefore, we expect weighted averages should be useful in calculating observables. Using (\ref{eq:weight}) to compute $\overline{E}$ and the bootstraps method to find the one standard deviation error, we obtain $\overline{E}=-224.175 \pm 0.037$ compared to the exact value, $E^{\rm exact}=-224.1896$ obtained from numerically summing the transfer matrices.
On the other hand, the unweighted average energy of the ensemble is $-223.987 \pm 0.024$.  Thus, the unweighted average is significantly biased while the weighted average shows no bias to within the statistical errors.  
The ground state energy for this disorder realization is $-226.861$ so at $1/\beta_0=5$ the system is quite close to its ground state.  For the free energy of the same realization we have $-\beta_0 F^{\rm exact} =1147.8697$ while the weighted average is $-\beta_0 \overline{F} = 1147.82 \pm 0.04$ and the unweighted average is  $1147.65 \pm 0.04$.  Again, the weighted estimator is not significantly biased while the unweighted estimator is biased.

\begin{figure}[t]
\includegraphics[width=0.45\textwidth]{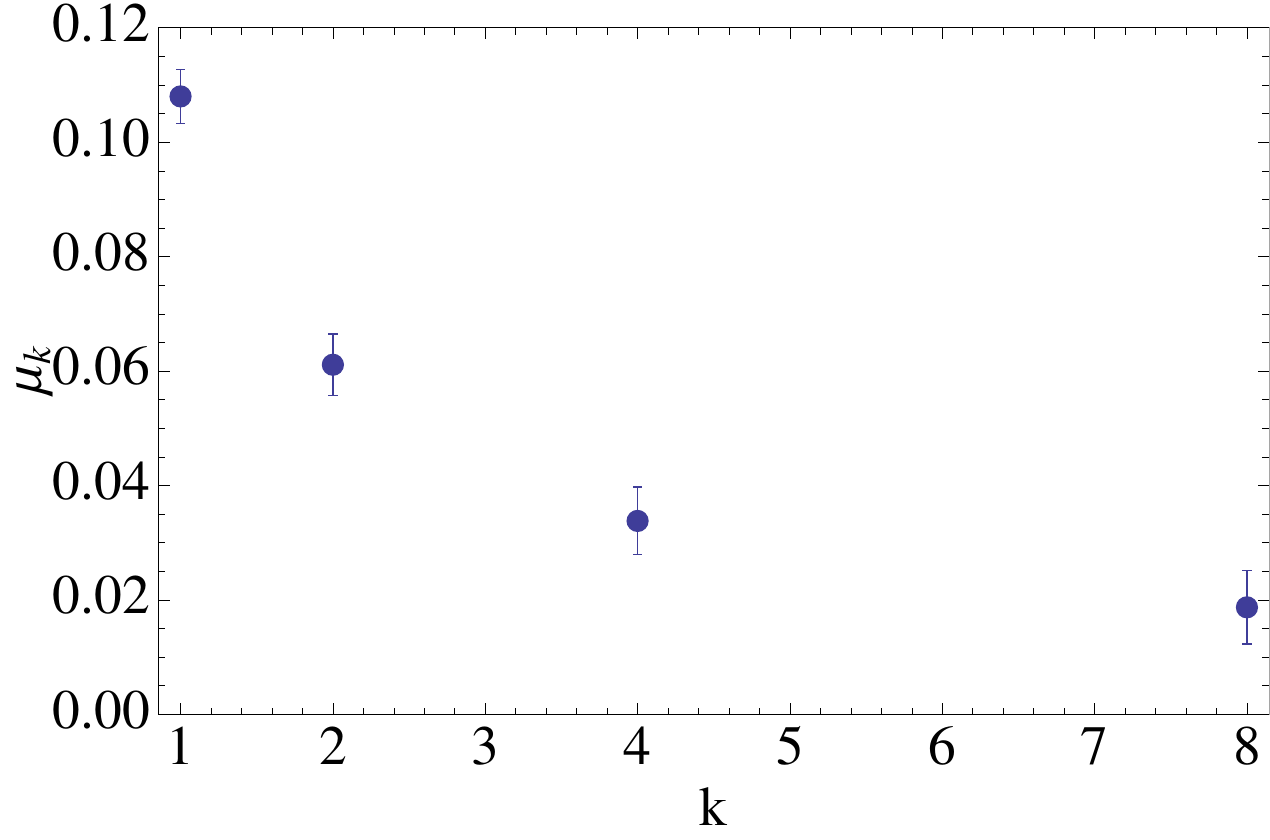}
\caption{The bias for the weighted energy estimator $\mu_k$ for ensemble size $k$ averaged over 18 disorder realizations for the 1D Ising spin glass.}
\label{fig:mukk}
\end{figure}

Next we consider the set of 18 disorder realizations and investigate how the bias decreases as the size of the ensemble of runs increases.  We divide the 200 independent runs for each disorder realization into $B$ blocks of size $k=200/B$ and then perform weighted averages for the energy within each of these blocks. Next we take the mean over the set of blocks and finally average these means over the 18 disorder realizations.  Let  $\mu_k$ be this disorder averaged bias,
\begin{equation} 
\mu_k= \frac{1}{18} \sum_{j=1}^{18} \left[(\frac{1}{B} \sum_{b=1}^B \overline{E}^b_j)-E^{\rm exact}_j\right]
\end{equation}
where $\overline{E}^b_j$ is the weighted energy estimator in block $b$ with block size $k=200/B$. We can similarly obtain the standard error of the mean from the $B$ blocks.  Figure \ref{fig:mukk} shows the bias $\mu_k$ as a function of block size $k$.  The case $k=1$ corresponds to unweighted averages.  By the time an ensemble of 8 runs is used, the bias is reduced by approximately a factor of 6.  The weighted energy estimator for the full ensemble of $M=200$ runs for each disorder realization yields a mean bias of $-0.0013 \pm 0.0069$ where the error estimate is obtained using the bootstraps method.  This weighted average displays no bias within the error bars.  Similar results hold for the free energy.  The $M=200$ weighted estimate of $-\beta \overline{F}$ averaged over the 18 disorder realizations deviates from the exact value by $0.0018 \pm 0.0083$, which is again indistinguishable from zero, while the unweighted free energy averaged over the same 18 realizations deviates from the exact answer by  $-0.1013  \pm 0.0077$.    

For the 3D Ising spin glass we studied 25 disorder realizations using \evo for system size 8$^3$.
The target population size is $R=2000$, $K+1=100$ temperatures are evenly spaced between $1/\beta_{K+1}=2$ and $1/\beta_0=0.2$ with $\theta=100$ Metropolis sweeps performed after every temperature step.    Each run requires 2$\times10^7$ Metropolis sweeps.  In addition to the energy and free energy, we considered the observable $\kk$, the probability that the overlap is less than 0.2, $\kk={\rm Prob}(|q|<0.2)$.  The overlap  is defined as $q=(1/N) \sum_{i=1}^N \sigma_i^{(1)}\sigma_i^{(2)}$ where the superscripts refer to two independent spin configurations of the system.   The probability distribution for the overlap reflects features of the free energy landscape.  When there is substantial weight for small $q$, there are two or more free energy minima that are widely separated but with comparable free energies.  For parallel tempering this situation is expected to lead to long equilibration times~\cite{Mac09a}.  Thus, it is interesting to look for correlations between the variance of $\beta \F$ and $\kk$. In the study of spin glasses, $\kk$ is important in distinguishing the competing pictures of the low temperature phase of the 3D Ising spin glass~\cite{MaZu99,PaYo00,KrMa00,KaYo02}.  Since $\kk$ involves the correlation between two independent replicas, we use pairs of independent runs to construct $q$. For each pair of runs, we compute $q$ by averaging over pairs of replicas, one from each run and then evaluate $\tilde{\kk}$ for that pair.  The appropriate weighted estimator $\overline{\kk}$ is 
\begin{equation}
\label{eq:pqs}
\overline{\kk} = \sum_{r=1}^M \tilde{\kk}_r \omega_r^\prime
\end{equation}
with
\begin{equation}
\label{eq:omegaprime}
\omega_r^\prime  = \frac{e^{-\beta( \F_r^{(1)} + \F_r^{(2)}) } }{\sum_{r=1}^M  e^{-\beta( \F_r^{(1)} + \F_r^{(2)}) }}
\end{equation} 
where the superscripts refer to the two of independent runs from which $\tilde{\kk}_r$ is constructed.  We did $M=50$ pairs of independent runs to compute $\overline{\Gamma}$.

\begin{figure}
\includegraphics[width=0.45\textwidth]{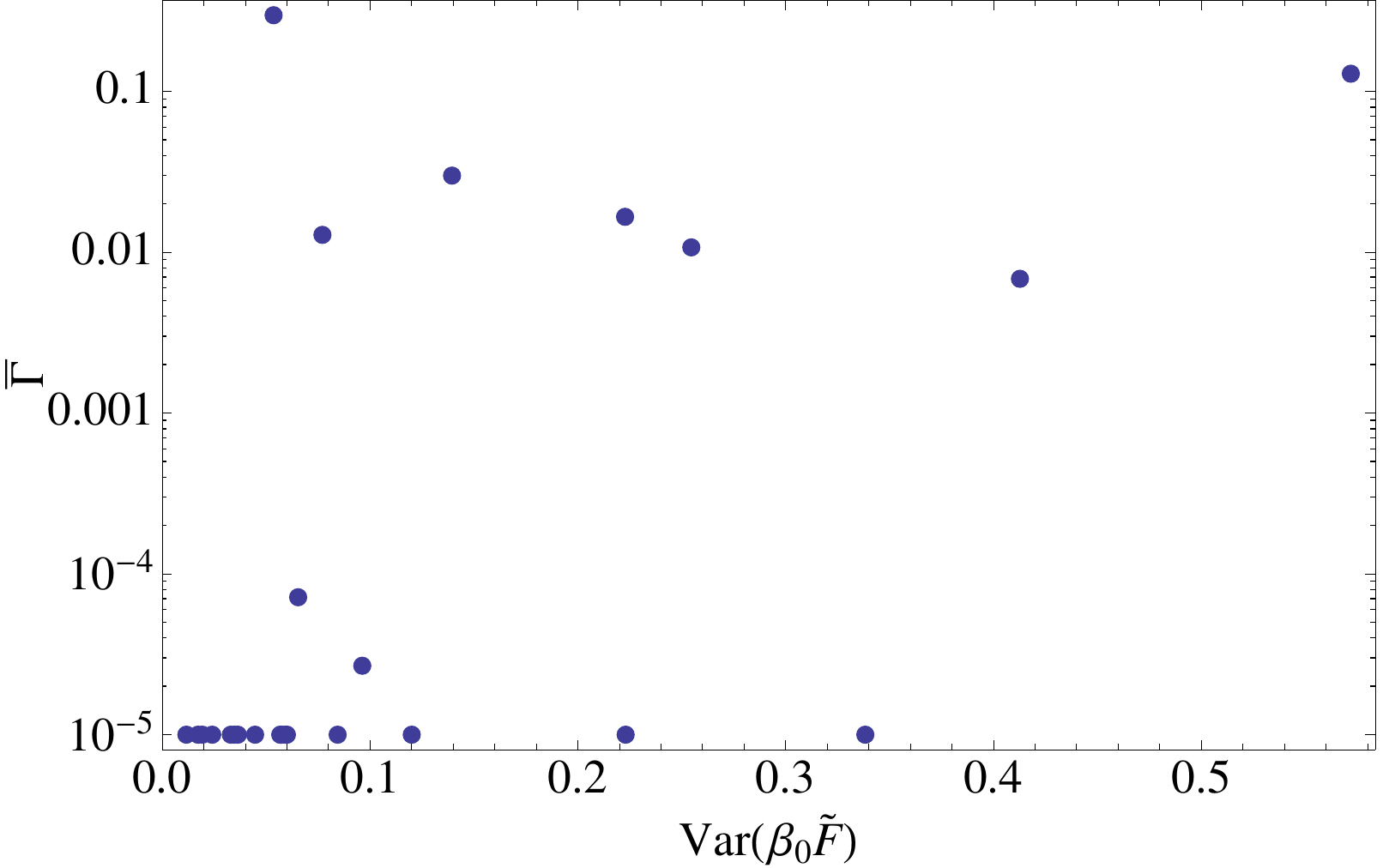}
\caption{The probability of a small overlap, $\overline{\kk}$ vs.\ variance of the free energy ${\rm Var}(\beta_0 \F)$, for a sample of 25 disorder realizations of the 3D Ising spin glass.}
\label{fig:varvspqs}
\end{figure}

Figure \ref{fig:varvspqs} shows a scatter plot of ${\rm Var}(\beta_0 \F)$  vs.\ $\overline{\kk}$ for the 25 disorder realizations.  The values of $\kk$ along the line $10^{-5}$ are estimated as zero by the algorithm on the basis of $10^5$ measurements of the overlap.  Most realizations have very small values of $\kk$, with only a few realizations dominating the disorder average.  The variance of the free energy estimator for the 25 realizations never exceeds 0.6.  There appears to be a correlation between a large variance of the free energy and a large value of $\kk$. Realizations with free energy variances less than 0.05 are all associated with very small values of the overlap near zero.  This correlation can be understood by realizing that $\kk>0$ implies that there are at least two free energy minima with comparable free energies.  If $R$ is not sufficiently large, the population of replicas may be dominated by one or another free energy minima in a single run and thus display a large variance in free energy from one run to another.

Of the 25 disorder realizations we now look more closely at the realization with the largest variance of $-\beta_0 \F$.
This `worst' realization has is ${\rm Var}(\beta_0 \F)=0.572$.  For this realization we obtained data from an ensemble of 125 runs (with $R=2000$, $K+1=100$ and $\theta=100$). The histogram of  $-\beta_0 (\F-\overline{F})$  is shown in Fig.\ 4 (left panel).  It should be noted that for the 3D simulations the highest temperature is not infinite, however, as we shall see below, a negligible contribution to ${\rm Var}(\beta_0 \F)$ arises from high temperatures. The mean of $-\beta_0 \F$ is 0.521 below the best estimate $-\beta_0 \overline{F}$, and this deviation is roughly equal to ${\rm Var}(\beta_0 \F)$ as expected.  The value of $\kk$ from these runs is $\overline{\kk}=0.129 \pm 0.011$ and the value of the energy is $\overline{E} =-847.294 \pm 0.015$. 
For this realization we also did simulations with a  larger target population size without changing $K$ or $\theta$.  The histogram of  $-\beta_0 (\F-\overline{F})$ for population size $R=10000$ is shown in  Fig.\ 4 (right panel).  For the larger population size ${\rm Var}(\beta_0 \F)=0.122$ so increasing the population size by a factor of 5 has reduced the variance by nearly the same factor.  The estimate for $\kk$ from this larger population size is $\overline{\kk}=0.145 \pm 0.005$, which is consistent with the results from the smaller population size.  The estimate for the energy is $\overline{E}=-847.317 \pm 0.009$, which is marginally consistent with the smaller run.  For comparison we simulated the same disorder realization using parallel tempering with two sets of 18 replicas each equally spaced in temperature with the same high and low temperatures.  The system was equilibrated for $10^6$ sweeps and data was collected for $10^6$ sweeps.  We ran this simulation 200 times to obtain means and errors.  Thus the total number of Metropolis sweeps for the parallel tempering runs was approximately $1.5 \times 10^{10}$ compared to approximately $1 \times 10^{10}$ for population annealing.
Parallel tempering runs yielded $\kk = 0.150 \pm 0.002$ and $E=-847.307 \pm 0.004$.  Results are marginally consistent for both $\kk$ and $E$. 
\begin{figure}
\label{fig:free3D}
\begin{tabular}{cc}
\includegraphics[width=0.45\textwidth]{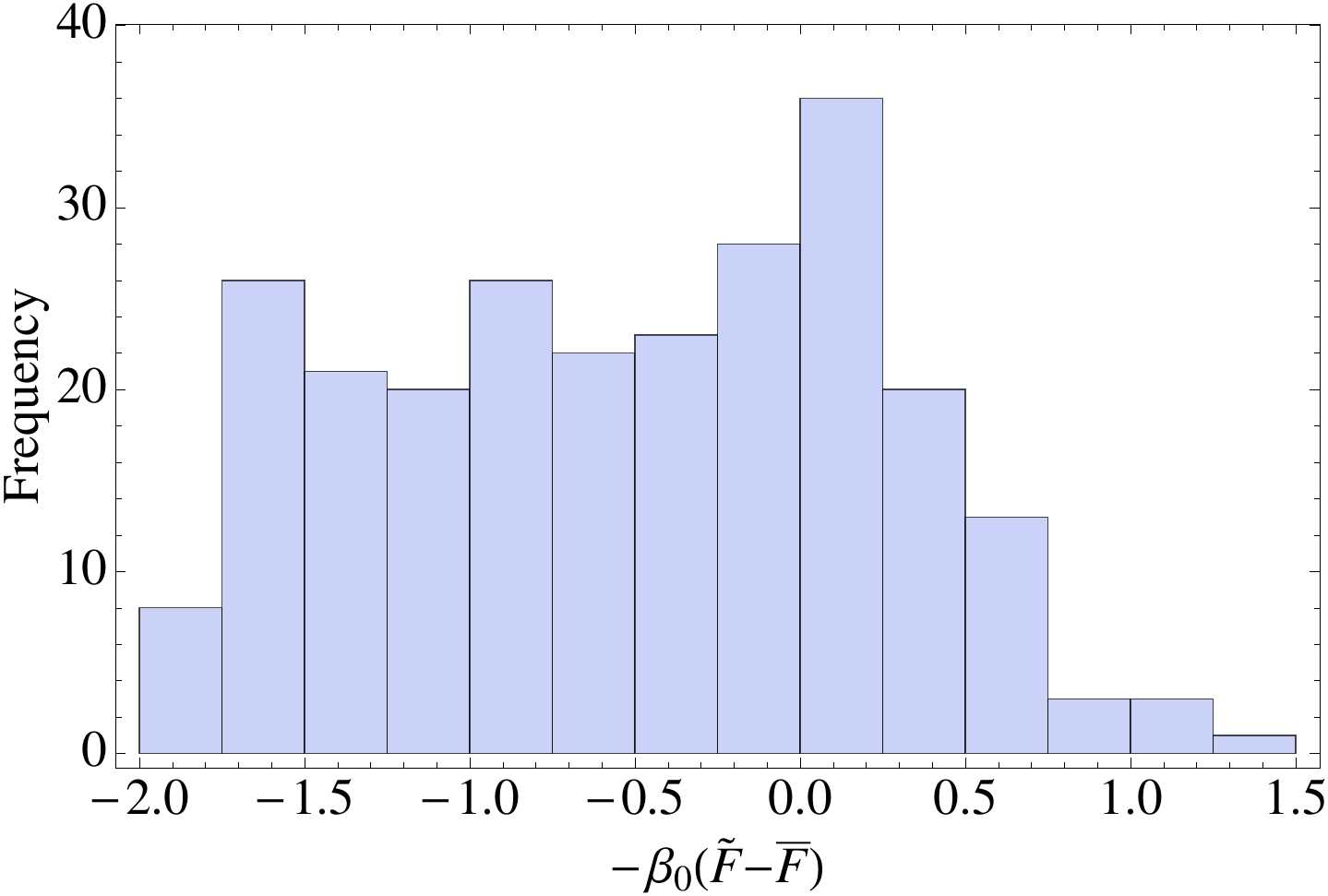}&
\includegraphics[width=0.45\textwidth]{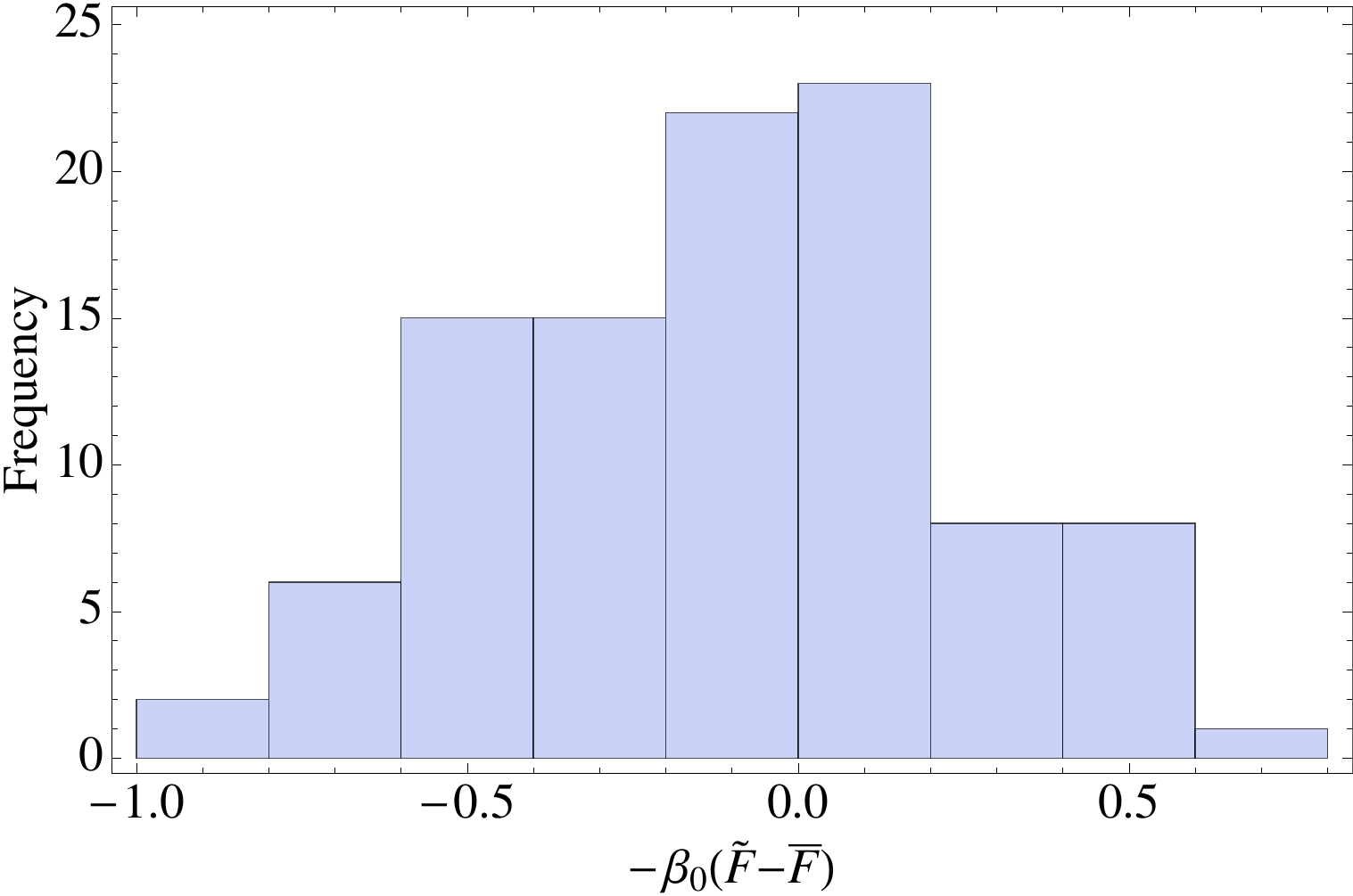}
\end{tabular}
\caption{The histogram of the dimensionless free energy deviation from its weighted average, $-\beta_0 (\F-\overline{F})$ for one realization of the 3D Ising spin glass.  In the left panel, the mean population size is $R=2000$ and in the right panel $R=10000$. }
\end{figure}

\begin{figure}
\includegraphics[width=0.75\textwidth]{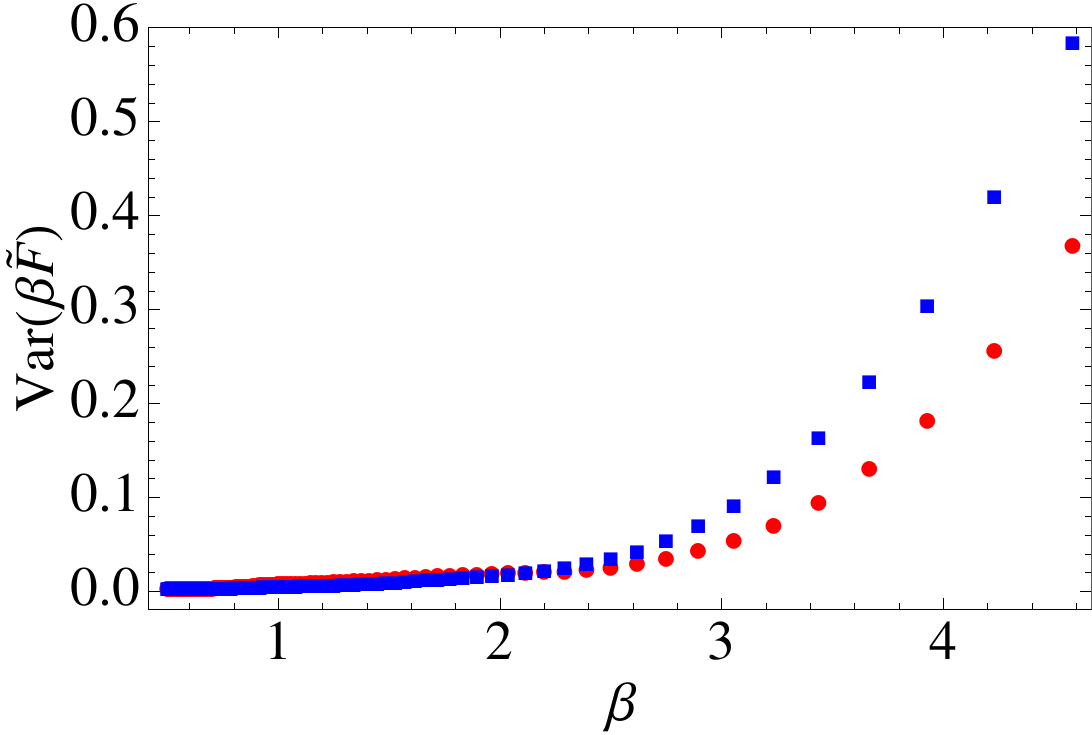}
\caption{(Color online)The  variance of the free energy estimator ${\rm Var}(\beta \F(\beta))$ vs.\ $\beta$ for one disorder realization for the 3D Ising spin glass. The upper (blue square) points show the result for a fixed population size ($R_\beta=2000$ for all $\beta$) while the lower (red circle) is for a population that grows for low temperature ($R_\beta=1000$ for $\beta<0.9$ and $R_\beta=3000$ for $\beta\geq0.9$).  Both simulations involve the same amount of computational work.}
\label{fig:vbf}
\end{figure}

Next we consider an initial attempt to improve the performance of the algorithm. The variance of the free energy estimator should be minimized to optimize the algorithm.  From (\ref{eq:sumQ}) we see that $-\beta \F(\beta)$ at each temperature is a sum of positive terms so that successively lower temperatures have successively larger variances.  Figure \ref{fig:vbf} shows ${\rm Var}(\beta \F)$ as a function of temperature for the `worst' disorder realizations with the largest variance discussed above.  The variance at each temperature is obtained from 50 iterations of the algorithm.  The upper points (blue online)  show the results from runs where the average population size is fixed, $R_\beta=2000$ for all $\beta$.  The variance increases slowly for small $\beta$ and then much more rapidly for large $\beta$.  The upward curvature of ${\rm Var}(\beta \F)$ suggests that more replicas should be used at low temperature.  The total computational work remains nearly fixed if $R_\beta=1000$ for $\beta<0.9$ and $R_\beta=3000$ for $\beta\geq0.9$.  The lower points (red online) in Fig.\ \ref{fig:vbf} show the results for this choice of target population size and demonstrates that, indeed, a significant reduction in ${\rm Var}(\beta \F)$ from 0.57 to 0.36 is achieved at the lowest temperature.  This attempt at optimizing the algorithm shows that there is room for improving its performance by careful choice of the parameters.

\section{Discussion}
Population annealing is a promising tool for measuring the free energy and other observables in spin glasses and perhaps other systems with rough free energy landscapes.  By using weighted averages over an ensemble of runs, biases inherent in a single run can be made small and high precision results can be obtained.    If the variance of the dimensionless free energy estimator ${\rm Var}(\beta \F(\beta))$ is less than about 0.5, high precision results can be obtained from a relatively small ensemble of runs.  This variance thus serves as measure of the convergence of the algorithm.

The method can be optimized by minimizing the variance of the free energy estimator.  In future work it would be important to optimize the method and study its efficiency and scaling with system size in comparison with well-established methods such as parallel tempering.

Compared to parallel tempering, population annealing is better suited to parallelization. Parallel tempering is optimized with a relatively small number of replicas.  Increasing the number of replicas in parallel tempering improves the acceptance rate of replica exchange moves but also lengthens the round trip distance between the warmest and coldest replicas.  Thus, the optimum number of replicas is relatively small in practice.  On the other hand, population annealing is monotonically improved by increasing the population size.  For example, for the $L=8$, 3D Ising spin glasses studied here population annealing makes effective use of thousands of replicas while parallel tempering is optimized with tens of replicas. Population annealing may find use in quickly equilibrating large systems using a very large population spread over many processors.
 
\begin{acknowledgments}
This work was supported in part from NSF grant DMR-0907235.  I thank Helmut Katzgraber, Burcu Yucesoy and Yukita Iba for helpful discussions.  
I thank the Santa Fe Institute for their hospitality and support.
\end{acknowledgments}

\end{document}